\documentclass[aps,prl,superscriptaddress,twocolumn,floatfix]{revtex4}

\usepackage{bm}
\usepackage{graphicx}
\usepackage{tabularx}

\def\dprat{F_{2d}/F_{2p}}
\def\nprat{F_{2n}/F_{2p}}

\begin{document}

\title{How well do we know the neutron structure function?}

\author{J. Arrington}
\affiliation{Physics Division, Argonne National Laboratory, Argonne, Illinois 60439, USA}

\author{J. G. Rubin}
\affiliation{Physics Division, Argonne National Laboratory, Argonne, Illinois 60439, USA}

\author{W. Melnitchouk}
\affiliation{Jefferson Lab, Newport News, Virginia 23606, USA}

\begin{abstract}

We present a detailed analysis of the uncertainty in the neutron
$F_{2n}$ structure function extracted from inclusive deuteron and proton
deep-inelastic scattering data.  The analysis includes experimental
uncertainties as well as uncertainties associated with the deuteron
wave function, nuclear smearing, and nucleon off-shell corrections.
Consistently accounting for the $Q^2$ dependence of the data and
calculations, and restricting the nuclear corrections to microscopic
models of the deuteron, we find significantly smaller uncertainty in
the extracted $F_{2n}/F_{2p}$ ratio than in previous analyses.
In addition to yielding an improved extraction of the neutron structure
function, this analysis also provides an important baseline that will
allow future, model-independent extractions of neutron structure to be
used to examine nuclear medium effects in the the deuteron.

\end{abstract}

\date{\today}

\maketitle

\begin{figure}[htb]
\centering
\includegraphics[width=0.45\textwidth]{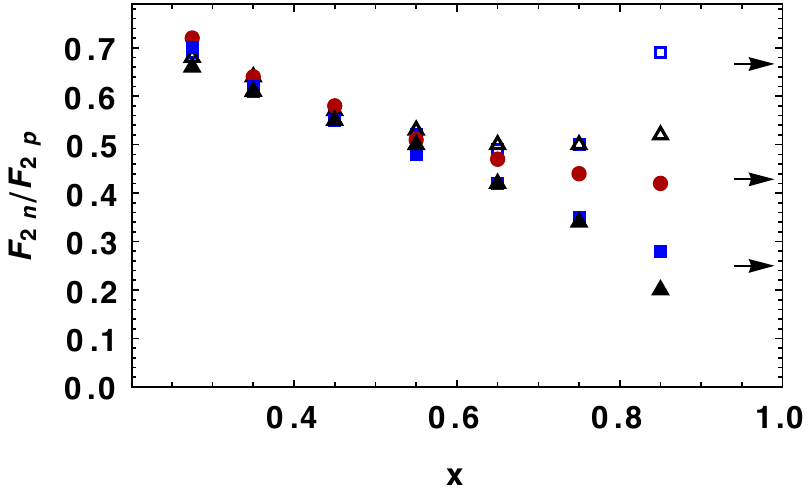}
\caption{(Color online) Previous extractions of $\nprat$, using
	microscopic deuteron calculations (filled symbols) or
	extrapolations of nuclear effects in heavier nuclei
	(open symbols):
	(black) triangles from Ref.~\cite{Whitlow:1991uw},
	(blue)  squares from Ref.~\cite{Bodek:1991iq},
	(red)   circles from Ref.~\cite{Melnitchouk96}.
	The arrows indicate theoretical predictions for the $x \to 1$
	limit (see text).\\}
\label{fig:oldRange}
\end{figure}

Because the free neutron is an experimentally impractical scattering
target, extracting its structure function, $F_{2n}$, requires cross
section data from inclusive deep-inelastic scattering (DIS) measurements
on proton and deuteron targets, together with a model describing the
smearing produced by the nuclear binding in the deuteron.  Previous 
extractions \cite{Whitlow:1991uw, Bodek:1991iq, Melnitchouk96} using
a variety of models of the deuteron bound state yielded a large range
of $\nprat$ values from the same proton and deuteron data, indicating
large theoretical uncertainties, particularly at high values of the
Bjorken variable $x$ --- see Fig.~\ref{fig:oldRange}.
In addition to limiting our ability to extract the neutron structure
function, the large spread of results, even among extractions including
only traditional nuclear effects such as Fermi motion and binding,
has made it difficult to identify a reliable baseline which could
be used to search for more ``exotic'' nuclear effects such as the
modification of the nucleon structure function in nuclei or
non-nucleonic degrees of freedom.

A more recent extraction of the neutron $F_{2n}$ structure function
\cite{Arrington09} showed that some of the variation in results
can be attributed to inconsistent treatment of kinematics of
the data and calculations.  In particular, it was vital to properly
account for the $Q^2$ dependence of the proton and deuteron data,
especially for $x \gtrsim 0.7$~\cite{Arrington09, Accardi10}.

Accurate information on the $\nprat$ ratio at large $x$ is essential
for a number of reasons.  These include constraining leading-twist
parton distribution functions (PDFs) in the region $x \gtrsim 0.7$,
which are an important input for QCD background calculations in searches
of physics beyond the standard model at the Tevatron and the LHC, and in
neutrino oscillation experiments.  The ratio $\nprat$ also provides
insight into the nonperturbative quark-gluon dynamics in the nucleon
\cite{Melnitchouk96, Holt10}.  There are several predictions for
$\nprat$ based on symmetry arguments that determine the dominant
contributions as $x \to 1$, ranging from $\nprat=2/3$ for exact
spin-flavor SU(6) symmetry, to 3/7 assuming helicity conservation
through hard gluon exchange~\cite{Farrar75}, to 1/4 when SU(6) symmetry
is broken through scalar diquark dominance~\cite{Feynman72, Close73}.

In this paper we extend the analysis of Ref.~\cite{Arrington09} by
performing a detailed study of the model dependence of the extraction
procedure, systematically assessing the uncertainties arising from the
deuteron wave function at short distances, nucleon off-shell effects,
and different nuclear smearing models used to compute the nuclear
corrections.  The goal is to determine the degree to which the $F_{2n}$
neutron structure function can be determined at large $x$, using
a consistent treatment of input data sets and realistic models of
nuclear effects, and employing a methodology that is transparent
and conceptually accessible.


Following Ref.~\cite{Arrington09}, we consider $F_{2d}/F_{2p}$ data
from SLAC, BCDMS, and NMC, as compiled in Ref.~\cite{Whitlow:1991uw}
for $3 < Q^2 < 230$~GeV$^2$, using the full $Q^2$ range to determine
the $Q^2$ dependence of the ratio.  The extraction of $\nprat$ itself
is limited to data in the range $8 < Q^2 < 32$~GeV$^2$, and the
results are interpolated to a fixed $Q^2 = Q_0^2 \equiv 16$~GeV$^2$.
The interpolation to $Q_0^2$ differs from the simple average over the
limited $Q^2$ range by at most 0.7\%, with a typical correction of 0.3\%.
Most previous extractions of $\nprat$ used of $F_{2d}/F_{2p}$ from the
analysis of Ref.~\cite{Whitlow:1991uw}, in which $Q^2$ varies from 4.7
to 23.6~GeV$^2$ over the $x$ range of the data.  However, the extractions
treat the $F_{2d}/F_{2p}$ ratios as though they were all at some average
$Q^2$ value.  Such extractions neglect the $Q^2$ dependence of the
nuclear effects, which have been found to be significant at large
$x$~\cite{Arrington09, Accardi10}. By interpolating all of the source
data to a common $Q_0^2$ and extracting $\nprat$ using different models
evaluated at the same scale, a more systematic and meaningful assessment
of the model dependence can be made.

The extraction of $\nprat$ proceeds by assuming that the deuteron
structure function $F_{2d}$ can be expressed as a sum of smeared
proton and neutron structure functions, and an additional correction,
$\delta F_{2d}$, that goes beyond the convolution approximation,
$F_{2d} = \overline F_{2p} + \overline F_{2n} + \delta F_{2d}$,
where $\overline F_{2N} = S_N\, F_{2N}$ is the smeared nucleon
structure function ($N=p, n$), and $S_N$ is the smearing ratio.
The term $\delta F_{2d}$ includes any corrections, such as relativistic
or nucleon off-shell corrections~\cite{Melnitchouk94,Kulagin2006126},
that cannot be expressed as a convolution of a smearing function and
the free nucleon structure function.  Parametrizing this correction
as a ratio to the total deuteron structure function,
$\Delta = \delta F_{2d} / F_{2d}$, the $\nprat$ ratio can then
be extracted using a modified smearing factor
$\widetilde{S}_N = S_N / (1-\Delta)$,
\begin{equation}
{ F_{2n} \over F_{2p} }
= {1 \over \widetilde{S}_n}
  \left( { F_{2d} \over F_{2p} } - \widetilde{S}_p F_{2p} \right).
\label{eq:F2np}
\end{equation}
The neutron to proton ratio can thus be extracted from $\dprat$
and $F_{2p}$ data, and a model of the smearing ratios $S_N$ and
the off-shell correction $\Delta$.

To standardize comparisons of the different calculations, all
extractions use the same values for $F_{2p}$ and $F_{2n}$ to
compute the smearing functions $\widetilde{S}_N(x)$.
We use the parametrization of the world's $F_{2p}$ data and the
extracted neutron structure function from Ref.~\cite{Arrington09}.
Each calculation yields a slightly different $F_{2n}$, however,
the impact of using this modified $F_{2n}$ value to update the
calculation of $S_n$ is small compared to the other uncertainties
($\nprat$ changes by less than 0.01 at $x=0.85$) \cite{Afnan03}.

The assessment of the model dependence of the extracted $\nprat$ ratio
ultimately depends on the choice of nuclear models used in the analysis.
The introduction of some degree of bias is therefore inevitable,
although we aim to make the selection criteria as objective as possible
by restricting ourselves to microscopic calculations involving
high-precision $NN$ potentials that give realistic descriptions of the
deuteron bound state.
Common features of the models surveyed in this analysis include the
use of realistic deuteron wave functions which account for nuclear
Fermi motion and binding, an exact treatment of finite-$Q^2$
kinematics, and allowance for possible nucleon off-shell corrections
in the deuteron~\cite{Melnitchouk96, Kahn09, Arrington09, Rinat03}.
We exclude models that involve extrapolations of nuclear medium
modifications observed in structure functions of heavy nuclei to
the deuteron~\cite{Frankfurt:1988nt, Whitlow:1991uw, Bodek:1991iq},
which typically invoke a very large nuclear density, contain no $Q^2$
dependence, and effectively assume $S_n=S_p$.  Note that this has
less impact than one would expect from Fig.~\ref{fig:oldRange}.
If one repeats these analyses taking into account the $Q^2$ dependence
of the $\dprat$ data, the ratios are below 0.5 at $x \gtrsim 0.6$
for all of the extractions.

\begin{figure}[htb]
\centering
\includegraphics[width=0.48\textwidth]{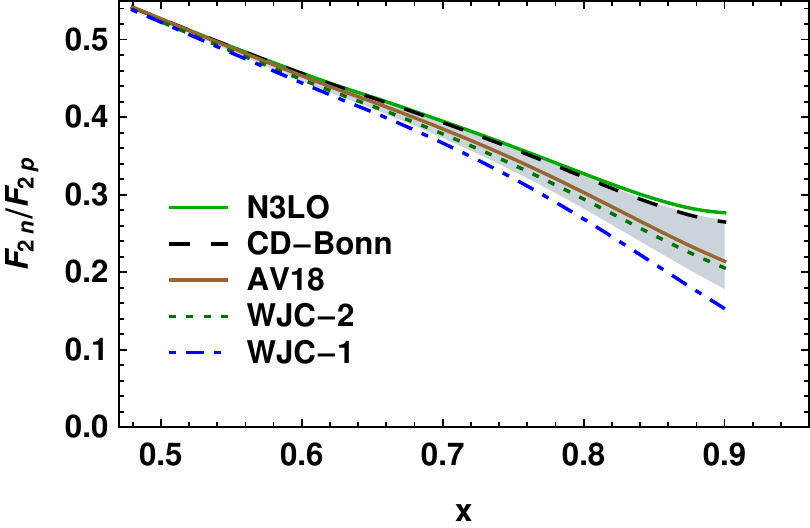}
\caption{(Color online) Neutron to proton structure function ratio
	$F_{2n}/F_{2p}$ calculated using various deuteron wave
	functions (see text), within the WBA smearing
	model~\cite{Kulagin2006126, Kahn09}.}
\label{fig:potentials}
\end{figure}

The dependence on the choice of deuteron wave functions is illustrated
in Fig.~\ref{fig:potentials}, which shows the results for $\nprat$
using different modern nonrelativistic (N3LO~\cite{entem03}, CD-Bonn 
\cite{Machleidt01}, AV18~\cite{wiringa95}) and relativistic (WJC-1,
WJC-2 \cite{gross07}) deuteron wave functions.  The smearing factors
for each of the calculations were computed using the weak binding
approximation (WBA) model~\cite{Kulagin2006126,Kahn09}, which is derived
by expanding the nucleon correlation function in the nucleus in powers
of the nucleon momentum ${\bm p}$ up to order ${\bm p}^2/M^2$, where $M$
is the nucleon mass.  The gray band represents the RMS spread of the
$\nprat$ ratios for the five wave functions considered, which is the
1$\sigma$ band for the wave function uncertainty if we treat each of
the $NN$ potentials on an equal footing.  The ratio $\nprat$ clearly
becomes increasingly sensitive to the choice of deuteron wave function
at larger $x$ values, reflecting the larger uncertainty in the $NN$
interaction at short distances or high nucleon momenta in the deuteron.

\begin{figure}[htb]
\centering
\includegraphics[width=0.45\textwidth,height=2.6in]{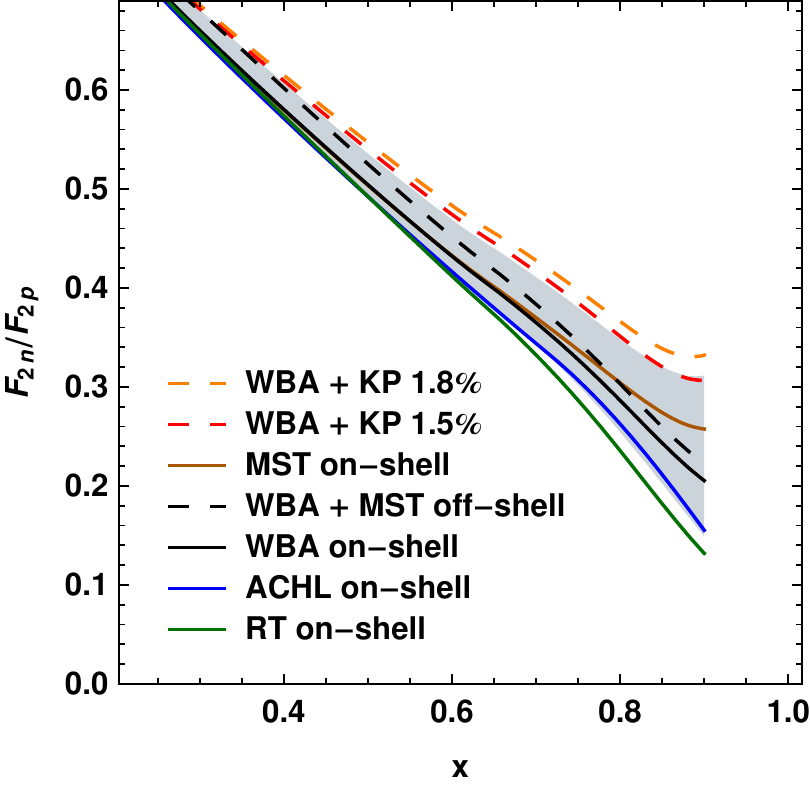}
\caption{(Color Online)
	The ratio $F_{2n}/F_{2p}$ calculated using different
	smearing models, taking the average of the $NN$ potentials
	from Fig.~\ref{fig:potentials}.  The solid curves are
	on-shell calculations:
	  WBA~\cite{Kulagin2006126,Kahn09},
	  relativistic MST~\cite{Melnitchouk94, Melnitchouk96},
	  the light-front ACHL~\cite{Arrington09}, and
	  the RT~\cite{Rinat03} models.
	The dashed curves show the WBA result with various
	off-shell prescriptions:
	  MST~\cite{Melnitchouk94}, and
	  KP~\cite{Kulagin2006126} with two different nucleon swelling
	parameters, 1.5\% and 1.8\% (see Ref.~\cite{Accardi11}).
	The ordering of the curves in the legend is based
	on the value of the extraction at $x=0.9$.}
\label{fig:models}
\end{figure}

The dependence of $\nprat$ on the model used for the smearing function
(or smearing factor) and off-shell prescription is illustrated in
Fig.~\ref{fig:models}.  The curves show the results for the given
calculation, averaged over the potentials shown in
Fig.~\ref{fig:potentials}.  For calculations where not all potentials
were available, the WBA result was used to extrapolate to the average
potential.  Note that some smearing calculations, such as those used
in Ref.~\cite{Whitlow:1991uw}, have been omitted as they represent
calculations similar to those included here, but with additional
numerical approximations.
The solid curves in Fig.~\ref{fig:models} are the results of smearing
calculations with on-shell nucleon structure functions, with their
spread indicating the uncertainty associated with the smearing function.
The WBA calculation (solid black curve) is a modern calculation that
makes minimal approximations, and is used as the baseline for showing
the results of calculations including nucleon off-shell corrections.
The four WBA results (solid black curve and the three dashed curves)
indicate the model dependence in the off-shell prescriptions.

To estimate the combined uncertainty, we take the RMS spread of all of
the extractions of $\nprat$ shown in Fig.~\ref{fig:models} at each $x$
value, indicated by the light gray band.  Similar uncertainties are
obtained if the model dependence of the smearing function and off-shell
contributions are extracted separately and combined in quadrature.
The uncertainty associated with the smearing function is essentially
negligible up to $x=0.6$, but is comparable to the off-shell corrections
for $x > 0.75$.

\begin{figure}[htb]
\centering
\includegraphics[width=0.48\textwidth]{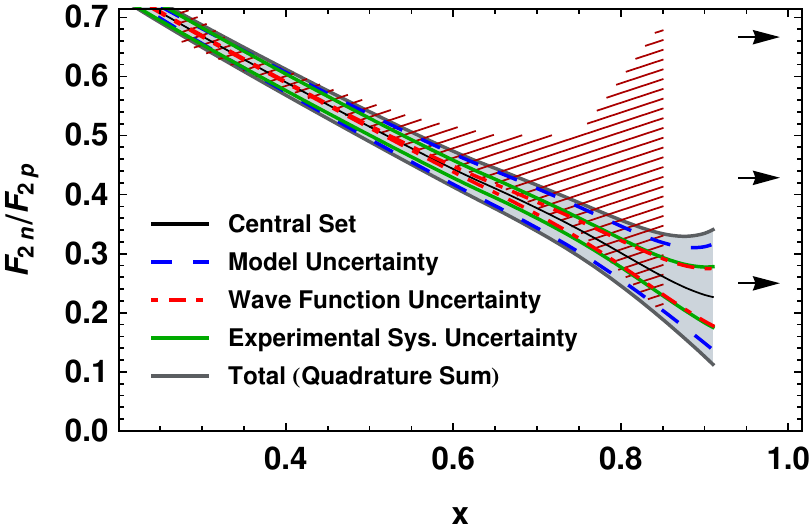}
\caption{$\nprat$ ratio together with the individual contributions
	to the systematic uncertainty and the quadrature sum.
	The red hatched region corresponds to the uncertainty range
	in Fig.~\ref{fig:oldRange}.}
\label{fig:combined}
\end{figure}

Figure~\ref{fig:combined} shows the combined uncertainty range (gray
band) compared to the range of results shown in Fig.~\ref{fig:oldRange}
(red hatched region).  The central result is taken as the global average
of the extracted $\nprat$ values obtained in Fig.~\ref{fig:models}.
The individual uncertainties associated with the experimental systematic
uncertainties (evaluated in Ref.~\cite{Arrington09}), the dependence
on the deuteron wave function, and the dependence on the smearing
function and off-shell effects (labeled ``Model Uncertainty'' in 
Fig.~\ref{fig:combined}), are shown separately, as well as the sum
of uncertainties added in quadrature.

Our analysis provides a significantly narrower range of results than
that evident in Fig.~\ref{fig:oldRange} for the full spectrum of models.
At $x=0.85$, for example, the range in Fig.~\ref{fig:oldRange} spans
$0.2 < \nprat < 0.7$, whereas the present analysis suggests a
1$\sigma$ range of $0.18 < \nprat < 0.32$.  The tighter bounds are
largely due to the exclusion of models involving extrapolation of
nuclear medium effects from heavy nuclei, and would appear to exclude
the SU(6) predictions of $\nprat \to 2/3$, while favoring the lower
estimates consistent with the partonic lower limit of $\nprat \to 1/4$.

\begin{figure}[htb]
\centering
\includegraphics[width=0.48\textwidth]{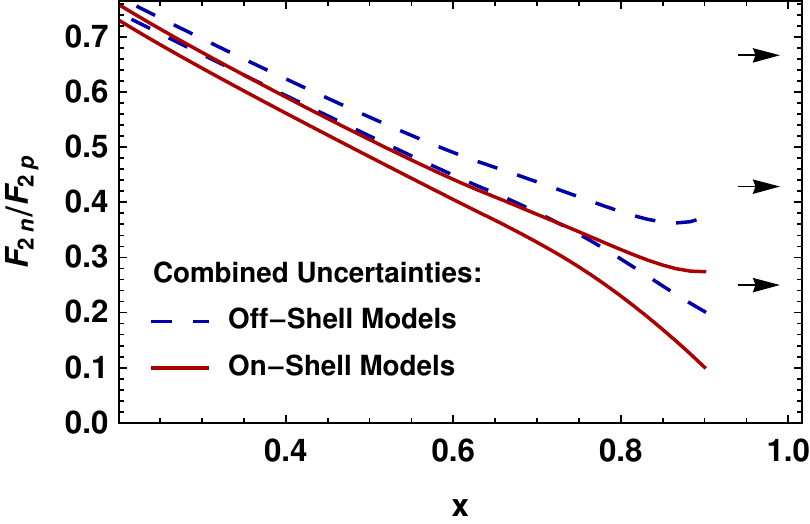}
\caption{$\nprat$ ranges for on-shell (solid) and off-shell (dashed)
	extractions.}
\label{fig:onoff}
\end{figure}

The dependence of the extraction on the choice of nuclear model is
further illustrated in Fig.~\ref{fig:onoff}, which shows the total
uncertainty bands corresponding to two different subsets of results.
The solid lines show the range obtained by taking only the on-shell
extractions in Fig.~\ref{fig:models}, which yield $0.16 < \nprat < 0.28$
at $x=0.85$.  While much of the range at large $x$ is below $\nprat=0.25$,
which in the parton model is forbidden by the requirement that PDFs
are positive, the results are not inconsistent with $\nprat>0.25$.
This range can be thought of as a baseline for effects beyond the
on-shell convolution approximation, and comparison of these results
to model-independent extractions of $\nprat$ can be used to isolate
off-shell contributions or other more exotic nuclear effects.

On the other hand, most modern quantitative analyses of nuclear
structure functions and the nuclear EMC effect require the inclusion
of some modification of the nucleon structure function in a nuclear
medium~\cite{Kulagin2006126,Smith:2003hu, Smith:2002ci,Cloet:2006bq}.
Restricting the set to only models that incorporate off-shell effects,
one obtains the dashed blue band in Fig.~\ref{fig:onoff}.  As expected,
this leads to a higher range for the neutron to proton ratio,
$0.25 < \nprat < 0.36$ at $x=0.85$.  Note that in both cases, the
experimental systematics and deuteron wave function dependence is
included in the bands.

\begin{figure}
\centering
\includegraphics[width=0.45\textwidth]{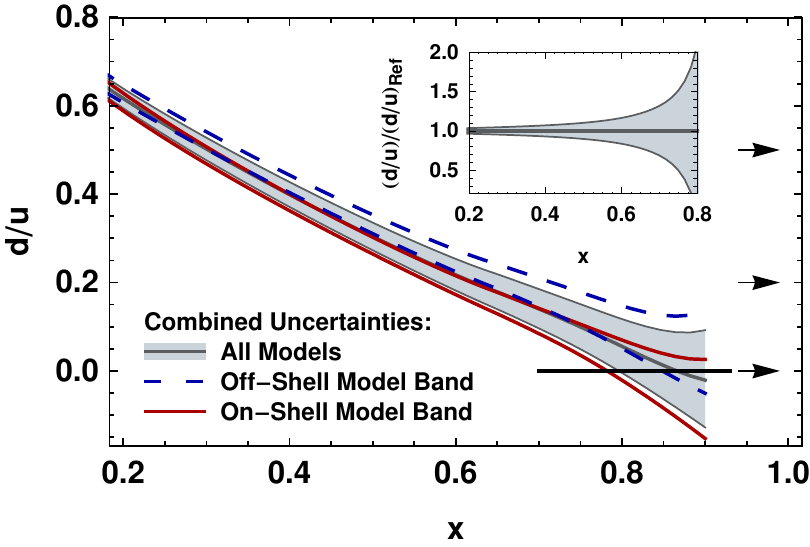}
\caption{Ratio $d/u$ extracted from $F_{2n}/F_{2p}$ at $Q^2=16$~GeV$^2$
	using Eq.~(\ref{eq:du}), with the gray band indicating the
	total uncertainties as shown in Fig.~\ref{fig:combined}, with
	the dashed (solid) curves representing the on-shell (off-shell)
	nuclear model results.  The inset shows the growing fractional
	uncertainty on $d/u$ as the ratio becomes smaller for $x \to 1$.}
\label{fig:uncertaintyComp}
\end{figure}

At leading order in the strong coupling constant, the nucleon structure
functions are given by the charge-squared weighted sum of the $u$ and
$d$ quark distributions.  In this approximation the extracted value
of $\nprat$ is directly related to the ratio of $d$ to $u$ quark
distributions,
\begin{equation}
{d \over u} = \frac{4 \nprat - 1}{4-\nprat} .
\label{eq:du}
\end{equation}
The resulting $d/u$ ratio is shown in Fig~\ref{fig:uncertaintyComp},
along with the fractional uncertainty (inset), for the full range
of models, as well as for the on-shell and off-shell models from
Fig.~\ref{fig:onoff}.  Such an extraction neglects higher-order
perturbative QCD corrections, target mass corrections, and higher twist
effects.  On the other hand, higher twists will cancel in this ratio,
unless they differ for the proton and neutron \cite{Alekhin:2003qq},
as will target mass corrections to a large extent, although some
residual prescription dependence survives at large values of $x$
\cite{brady11}.  Therefore, even though Eq.~(\ref{eq:du}) is
approximate, the results in Fig.~\ref{fig:uncertaintyComp} do serve to
illustrate how uncertainties in $\nprat$ propagate to the $d/u$ ratio.

The absolute uncertainty on the $d$ quark distribution is small
compared to the overall size of the $u$ quark distribution
($< 10\%$ for all $x$ values shown).  In contrast, the {\it fractional}
uncertainty on the $d$ quark PDF is large, and will yield significant
uncertainties on quantities sensitive to small relative uncertainties
in $d$, such as the PDF inputs for cross section calculations in
high-energy collisions.  Although the error band in
Fig.~\ref{fig:uncertaintyComp} includes a significant region
with $d < 0$ for $x > 0.8$, corresponding to $\nprat < 0.25$,
the 1$\sigma$ bands are never inconsistent with a positive PDF.

While the extracted $d/u$ ratio in Fig.~\ref{fig:uncertaintyComp} is
illustrative of the reduced uncertainty from the restricted range of
nuclear corrections considered here, the primary goal of the present
work is an extraction of the total $F_{2n}$ rather than a separation
of the PDFs from target mass and higher twist contributions.
A complementary study which quantified the effects of nuclear
corrections on PDFs within a global QCD analysis was recently
performed by the CJ Collaboration \cite{Accardi11}.  The largest
impact of the nuclear effects, which were computed using the WBA
smearing function with nucleon off-shell corrections from the modified
KP model (see Fig.~\ref{fig:models}), was found for the $d$ quark PDF
at large $x$.  The uncertainties in the resulting $d/u$ ratio were
similar to those in Fig.~\ref{fig:uncertaintyComp}, although with a
somewhat larger spread.  
In particular, the range in Ref.~\cite{Accardi11} for the neutron to
proton ratio at $x=0.85$ was found to be $0.32 < \nprat < 0.50$ at
$Q^2 = 16$~GeV$^2$, where the upper and lower limits were obtained
using models with the minimum and maximum nuclear corrections,
respectively.  This is a larger range than found here, although
it represents the full range of results, rather than an estimated
1$\sigma$ error band.  Taking a linear sum of all uncertainties from
Fig.~\ref{fig:combined} yields the range $0.04 < \nprat < 0.42$.
The upper limit is in reasonable agreement with the CJ global fit,
while the lower limit is significantly smaller.  This is because PDFs
in global QCD analyses are constrained to be positive {\it a priori},
forcing the overall $d/u$ bands (and consequently $F_{2n}/F_{2p}$)
to lie higher than those in Fig.~\ref{fig:uncertaintyComp}.

In summary, we have extracted the $\nprat$ structure function ratio
from a range of models of the deuteron structure, and estimated the
uncertainties associated with the choice of nuclear smearing model,
deuteron wave function, nucleon off-shell corrections, and experimental
uncertainties.  Restricting the analysis to models based on microscopic      
calculations of deuteron structure, and excluding those that rely
on extrapolations from heavier nuclei, we find a range of results
that is significantly smaller than in some previous extractions.
The general consistency between different microscopic models evaluated
here provides a reliable baseline for the neutron structure assuming
only traditional nuclear effects, so that future model-independent
extractions of the neutron structure function~\cite{e1210102, e1210103}
can both improve our knowledge of the neutron structure and have the
potential to provide signatures for more exotic nuclear effects.


This work was supported by the US DOE under contracts DE-AC02-06CH11357
and DE-AC05-06OR23177, under which Jefferson Science Associates, LLC
operates Jefferson Lab.  We thank S.~Kulagin and A.~Rinat for providing
calculations and A.~Accardi and J.~Owens for helpful discussions.

\bibliographystyle{apsrev}
\bibliography{neutron}

\vfill\eject
\end{document}